\documentclass[reprint,
twocolumn,
preprintnumbers,
bibnotes,
pra,
lengthcheck
]{revtex4-2}
\usepackage{amsmath}
\usepackage{eurosym}
\usepackage{graphicx}
\usepackage{dcolumn}
\usepackage{bm}
\usepackage{dsfont}
\usepackage{newtxtext,newtxmath}
\usepackage[colorlinks,urlcolor=blue,linkcolor=blue,citecolor=blue,anchorcolor=blue]{hyperref}
\usepackage{color}
\usepackage[ruled,lined]{algorithm2e}

\begin{document}
	
	\title{Generation of strongly localized skin solitons in non-Hermitian
		waveguide arrays with the Kerr effect}
	\author{Chong Hou$^1$}
	\author{Boris A. Malomed$^{2,3}$}
	\author{Qin Zhou$^{1,4}$}
	\email{qinzhou@whu.edu.cn}
	\affiliation{$^1$Research Group of Nonlinear Optical Science and Quantum
		Technology, School of Microelectronics, Wuhan Textile University, Wuhan
		430200, China}
	\affiliation{$^2$Department of Physical Electronics, School of Electrical and Computer
		Engineering, Faculty of Engineering, Tel Aviv University, P.O.B. 39040, Ramat Aviv,
		Tel Aviv, Israel}
	\affiliation{
		$^3$Instituto de Alta Investigación, Universidad de Tarapacá,
		Casilla 7D, Arica, Chile}
	\affiliation{$^4$State Key Laboratory of New Textile Materials
		and Advanced Processing, Wuhan Textile University, Wuhan 430200, China}
	
	
	\begin{abstract}
		We address two distinct nonlinear propagation problems in nonlinear optical
waveguide arrays (WGAs) with non-reciprocal (non-Hermitian) couplings.
First, we investigate the light propagation launched by initial excitations
of two different types. The single-channel excitation creates stable
solitons supported by the interplay of the Kerr nonlinearity and
non-Hermitian skin effect (NHSE). In this case, we derive, by means of the
symbolic-regression method, an analytical formula defining the soliton
existence boundary. For the broad-pulse excitation, we produce perturbed
soliton solutions analytically in the continuum approximation, which is
accurately corroborated by numerical results. We thus conclude that NHSE
accelerates the propagation of the broad soliton towards the boundary,
ultimately causing tight localization at the edge, which is a hallmark of
the NHSE in the continuum limit. Second, we identify stationary solitons in
the system -- specifically, nonlinear bulk modes in the Hermitian regime and
near-edge skin solitons in the non-Hermitian one. The nonlinear bulk modes
are compressed toward the edge of the WGA under the action of the
non-reciprocality, which is the nonlinear extension of NHSE.
	\end{abstract}
	
	\maketitle
	
	\affiliation{$^1$Research Group of Nonlinear Optical Science and Quantum Technology, School of Microelectronics, Wuhan Textile University, Wuhan 430200, China}
	\affiliation{$^2$Department of Physical Electronics, School of Electrical and Computer Engineering, Faculty of Engineering, Tel Aviv University, P.O.B. 39040, Ramat Aviv, Tel Aviv, Israel}
	\affiliation{
		$^3$Instituto de Alta Investigación, Universidad de Tarapacá, Casilla 7D, Arica, Chile}
	\affiliation{$^4$State Key Laboratory of New Textile Materials and Advanced Processing, Wuhan Textile University, Wuhan 430200, China}
	
	
	\section{Introduction}
	
	\label{Sec1}
	
	Waveguide arrays provide an ideal platform for the realization of optical manipulations in various forms. Studies of waveguide arrays date back to 1965, when the effect of the evanescent field coupling between adjacent waveguides was discovered. {This effect induces} discrete diffraction and energy exchange of light in tightly coupled waveguiding arrays (WGAs) \cite{jones_coupling_1965}. In 1998, discrete solitons in Kerr-nonlinear WGAs were created. It was found that, under the single-channel excitation, the optical output in WGAs gradually exhibits localization with the increase of the pump power: high power induces strong spatial self-localization, suppressing diffraction and reducing losses. The WGA dynamics is adequately modeled by the discrete nonlinear Schr\"{o}dinger (DNLS) equation \cite{eisenberg1998discrete}.
	
	Studies of WGAs attract much interest for various reasons. On the one hand, driven by technological advancements, femtosecond laser direct-writing technology has enabled the fabrication of complex WGAs with arbitrary spatial structures in transparent media, such as quartz and lithium niobate, providing a solid material foundation for related studies \cite{kevrekidis_discrete_2009, szameit2010discrete, gorbach_topological_2025}. On the other hand, recent breakthroughs in the non-Bloch band theory offer new perspectives for WGA systems \cite{yok19,Ozawa2019}. In this vein, combining topology and non-Hermiticity has unveiled a wealth of novel phenomena \cite{price_roadmap_2022, yan_advances_2023, song_artificial_2025}, such as the non-Hermitian skin effect (NHSE) \cite{zhang_review_2022}, non-reciprocal wave propagation \cite{wei20,eichelkraut2013mobility, Longhi2016nonher, wanjura_topological_2020, hou_controlling_2025}, exceptional points \cite{Zhen2015, Cer19, Ali2019}, photonic Floquet topological insulators \cite{rechtsman_photonic_2013} and PT (parity-time) symmetric waveguides \cite{xia_nonlinear_2021}. 
   {The study of nonlinear lattice dynamics provides} a theoretical foundation for light propagation in WGA systems \cite{aleshkevich_eigenvalue_2004, malomed_spatiotemporal_2005, kartashov_surface_2006, kartashov_gap_2008, kartashov_solitons_2011, malomed_multidimensional_2016, kartashov_frontiers_2019}. In this context, WGAs serve as a versatile platform for investigation of the combination of non-Hermitian, nonlinear, and dispersive effects, revealing a variety of topological self-trapped {modes} \cite{smirnova_nonlinear_2020,parto_non_hermitian_2020,price_roadmap_2022,zhang_second_2023,ezawa_nonlinearityinduced_2021,komis_skin_2023} and facilitating the realization of new effects and applications, such as the nonlinear Thouless pump \cite{jurgensen_quantized_2021,walter_quantization_2023}, topological edge states \cite{hadad_selfinduced_2018}, bulk solitons emerging from topological bands \cite{mukherjee_observation_2020}, and topological lasers \cite{smirnova_nonlinear_2020,zykin_topological_2021,zhong_theory_2021,leefmans_topological_2024}.
	
	The NHSE, characterized by the exponential localization of bulk-state wavefunctions at {the system boundaries}, has gained much attention \cite{Yao18,FSong19,Zhang2022,gohsrich_nonhermitian_2025}. It originates from non-Hermitian mechanisms based on asymmetric hopping or gain and loss, causing a dramatic reorganization of the energy spectrum under open boundary conditions. This phenomenon renders the traditional Bloch-band theory inapplicable, necessitating the introduction of generalized Brillouin zones \cite{Yao18,Hou22,hou_topological_2023,hou_non_hermitian_2024,zhang_review_2022, Zhang2022,li_dual_2024,li_exact_2025,chen_nonhermitian_2025,wang_observation_2025}. In WGAs, the NHSE facilitates unidirectional light transport and localized accumulation, often accompanied by gain/loss effects \cite{hou_controlling_2025,wang_nonlinear_2025}. However, detailed analysis of mechanisms underlying these wave phenomena---in particular, the interplay of NHSE with nonlinearity---remains an open question. Recently, one study investigated these phenomena within the framework of Hatano-Nelson (HN) lattices \cite{komis_skin_2023}. By examining the interplay between nonlinear self-focusing and non-Hermitian phase effects under single-channel excitation, they identified skin solitons characterized by a power threshold and marked spatial asymmetry. Another work \cite{many_manda_skin_2024} demonstrated the manifestation of the NHSE in a nonlinear HN system using perturbation theory and numerical calculations, thereby investigating the existence, morphology, and stability of nonlinear skin modes. Building upon these pioneering works, the present study aims to investigate similar phenomena within an extended nonlinear non-Hermitian HN model.
	
	Distinctly from previous studies, we here investigate optical transport driven by the combined action of nonlinearity and non-Hermiticity under single-site or broad-pulse excitation. In these two cases, the ensuing dynamics follows fundamentally different scenarios. In the former case, leveraging the Participation Ratio (PR) metric {as proposed in Ref.~\cite{komis_skin_2023}}, we employ the symbolic-regression technique to establish an analytical relation between the NHSE and nonlinearity, thereby quantifying regimes of soliton formation. In the latter case, we establish, for the first time, a correspondence between the discrete HN model and the continuum NLS equation. Using perturbation theory, we conclude that the NHSE triggers accelerated motion of the wave packet toward the boundary, eventually leading to pinning at the WGA edge. Furthermore, unlike Ref.~\cite{many_manda_skin_2024}, our analysis of static nonlinear modes reveals distinct classes of nonlinear bulk modes and skin solitons. We show that, under the action of the NHSE, the bulk modes preferentially localize toward the boundary, {evolving} into skin solitons---a behavior that mirrors the linear NHSE and represents its nonlinear counterpart, a phenomenon that has not been previously reported. In addition, we provide an analytical solution for the eigenvalues and eigenstates of the bulk modes.
	
	The non-Hermitian HN model was proposed in 1996 to describe a superconducting system \cite{Hatano96}; {and it was} later experimentally emulated in optics \cite{liu_complex_2022}. Analytical solutions of this model reveal that its eigenvalues are real-valued, zero modes do not exist in it, and its eigenstates (``skin modes") exhibit the NHSE \cite{guo21,hou_non_hermitian_2024}. The HN model was extended to its nonlinear version by adding the Kerr term to the lattice equation. In optics, the nonlinear HN model can be implemented as an appropriate WGA \cite{eisenberg1998discrete,ma2017quantum,zhong_theory_2021,hou_controlling_2025}. A recent work implementing the HN model in Bose-Einstein condensates {has} demonstrated that Raman-engineered spin-orbit currents generate an effective imaginary potential, enabling collective non-reciprocity and centroid self-acceleration \cite{tao_imaginary_2026}. {Related experiments were also performed in cold-atom platforms \cite{liang_dynamic_2022} and driven-dissipative cavity arrays \cite{wanjura_topological_2020}.} In the Hermitian limit, the HN model reduces to the DNLS equation, which supports currently known lattice solitons \cite{malomed_soliton_1996,eisenberg1998discrete,lederer2008discrete,vicencio_discrete_2009,kevrekidis_discrete_2009,ferreira_dissipative_2022,hou_lump_2025,chen_versatile_2017,chen_modulation_2022}.
	
	The subsequent presentation is organized as follows. Section \ref{Sec2} introduces the nonlinear extension of the non-Hermitian HN lattice model and its experimental implementation. Section \ref{Sec3} examines the optical transmission governed by the interplay of the NHSE and nonlinearity in two scenarios, \textit{viz}., with the single-site or broad-pulse excitation, the latter one addressed in both numerical and analytical forms. Section \ref{Sec4} deals with stable stationary solutions for nonlinear bulk modes and skin solitons. Section \ref{Sec5} concludes this work.

	\section{The nonlinear Hatano-Nelson system}
	
	\label{Sec2}
	
\begin{figure}[th]
		\begin{center}
			\includegraphics[width=8.3cm]{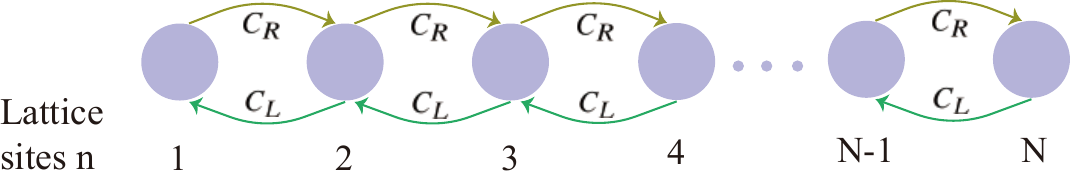}
		\end{center}
		\caption{
			The schematic of the nonreciprocal nonlinear non-Hermitian HN lattice composed of~$N$ sites (active optical waveguides), which corresponds to Eq.~(\protect\ref{eq1}), where $C_{L}$ and~$C_{R}$~are the nonreciprocal hopping parameter, with the non-Hermiticity strength defined as $h\equiv C_{R}-C_{L}$. Note that onsite potential energy induced by Kerr nonlinearity is not explicitly visualized in the diagram, as this effect emerges spontaneously during light propagation. }
		\label{fig1}
	\end{figure}

	Our starting point is a one-dimensional lattice composed of $N$~waveguides with non-reciprocal coupling between them, which can be derived in the framework of the coupled-mode theory \cite{eisenberg1998discrete, christodoulides2003discretizing, szameit2008long,lederer2008discrete,skryabin2021waveguide,mohammadi2024tunable,yang2025programmable}. Under the paraxial approximation, the phenomenological evolution of the optical fields is governed by the nonlinear HN model, written as a system of normalized coupled equations for field amplitudes $U_{n}$ at the $n$-th lattice site ($n=1,...,N$, see Fig.~\ref{fig1}):
	\begin{equation}
		i\frac{d}{dz}U_{n}=C_{L}U_{n+1}+C_{R}U_{n-1}+\sigma |U_{n}|^{2}U_{n}.  \label{eq1}
	\end{equation}
	Here, $z$ is the propagation distance, and $C_{L}$ and $C_{R}$ are nonreciprocal hopping parameters. The system reduces to the standard (Hermitian) DNLS equation in the reciprocal case where $C_{L}=C_{R}^{\ast }$ (with $\ast$ denoting the complex conjugate). The real parameter $\sigma$ represents the Kerr nonlinearity coefficient. In this work, we focus on the case with purely real asymmetric hopping parameters $C_{L,R}$. Consequently, the strength of the  {non-Hermiticity} in the NH model is characterized by the real parameter,
	\begin{equation}
		h=C_{R}-C_{L}.  \label{h}
	\end{equation}
	We adopt the convention that $\sigma<0$ and $\sigma>0$ correspond respectively {to} the self-focusing and defocusing nonlinearity. As usual, it is sufficient to consider the case of $\sigma >0$, as the sign of $\sigma $ may be inverted by means of the staggering transform \cite{kevrekidis_discrete_2009}: $U_{n}\equiv (-1)^{n}\tilde{U}_{n}^{\ast }$. Equation~(\ref{eq1}) may be naturally supplemented by boundary conditions of two different types: periodic, defined by $U_{0}\equiv U_{N}$, $U_{N+1}\equiv U_{1}$, or fixed (Dirichlet), imposed by setting~$U_{0}=U_{N+1}\equiv 0$. In this work, aiming to study skin modes attached to the {system boundaries}, we focus on the latter type.
	
	Femtosecond laser writing technology can be employed to fabricate such optical WGAs in photorefractive crystals. The coupling strength between adjacent waveguides in the array is determined by the spacing between them, with a smaller spacing corresponding to stronger coupling \cite{jones_coupling_1965,szameit2010discrete,zeuner_observation_2015}. {The non-reciprocity of the system, manifested by $C_{L}\neq C_{R}$, is achieved through the formation of Aharonov–Bohm photonic cages \cite{longhi_aharonovbohm_2014,gou_tunable_2020}, which result from the combination of a Peierls phase and active lossy sites under uniform external pumping \cite{liu_complex_2022,Longhi2016nonher}. Indeed, the presence of the pump and loss is consistent with the non-conservative character of the non-Hermitian system.}
	
	\section{Lattice solitons in WGAs (waveguiding arrays)}
	
	\label{Sec3}
	
	{\color{black} This section investigates the WGA dynamics under two excitation schemes---single-site and broad-pulse---with a focus on wave-packet localization driven by the interplay between nonlinearity and the NHSE. Due to the structural analogy between these WGAs and condensed-matter lattices, we refer to such {self-trapped, localized wave packets} as \textit{lattice solitons}. While the fundamental localization mechanism in the Hermitian case---balancing nonlinearity against discrete diffraction---aligns with landmark observations of discrete spatial optical solitons \cite{eisenberg1998discrete}, the states investigated here differ significantly: these are dissipative structures inherent in discrete non-Hermitian systems, and wave-packet stability is no longer governed solely by the power-diffraction balance, but is affected by a more complex dynamical equilibrium involving nonreciprocal hopping. }
	
	Regarding the single-site excitation regime, while the term ``localized waves" might be technically more precise, given the absence of analytical solutions in this case, we maintain terminological consistency with the current literature by classifying these states as lattice solitons. In this framework, a wave is characterized as a lattice soliton if it exhibits sustained self-focusing over a significant propagation distance, with the corresponding PR approaching unity or remaining sufficiently low. This operational definition provides a robust foundation for elucidating the interplay between the nonlinearity and non-Hermiticity in our analysis.
	
	\subsection{Lattice solitons produced by the single-site excitation}
	
	The single-channel (single-site) excitation for Eq.~(\ref{eq1}) is defined as
	\begin{equation}
		U_{n}(z=0)=A\delta _{n,n^{\prime }}.  \label{input}
	\end{equation}
	Here $A$ is the input amplitude and~$\delta _{n,n^{\prime }}$ is~the {Kronecker delta} for the input applied at the site with coordinate $n^{\prime }$. Obviously, the input amplitude $A$ and nonlinearity coefficient $\sigma $ exert equivalent influence on the system's dynamics, determined by {the} product $\sigma A^{2}$. The equivalence is demonstrated by substituting the scaled field variable $U_{n}=AV_{n}$ in Eq.~(\ref{eq1}), resulting in
	\begin{equation}
		i\frac{d}{dz}V_{n}=C_{L}V_{n+1}+C_{R}V_{n-1}+\alpha |V_{n}|^{2}V_{n},
		\label{V}
	\end{equation}
	with the~\textit{optical nonlinearity coefficient} (ONC) {defined as} $\alpha \equiv \sigma A^{2}$, while the input condition (\ref{input}) is replaced by $V_{n}(z=0)=\delta _{n,n^{\prime }}$.
	
	\begin{figure}[th]
		\begin{center}
			\includegraphics[width=8.7cm]{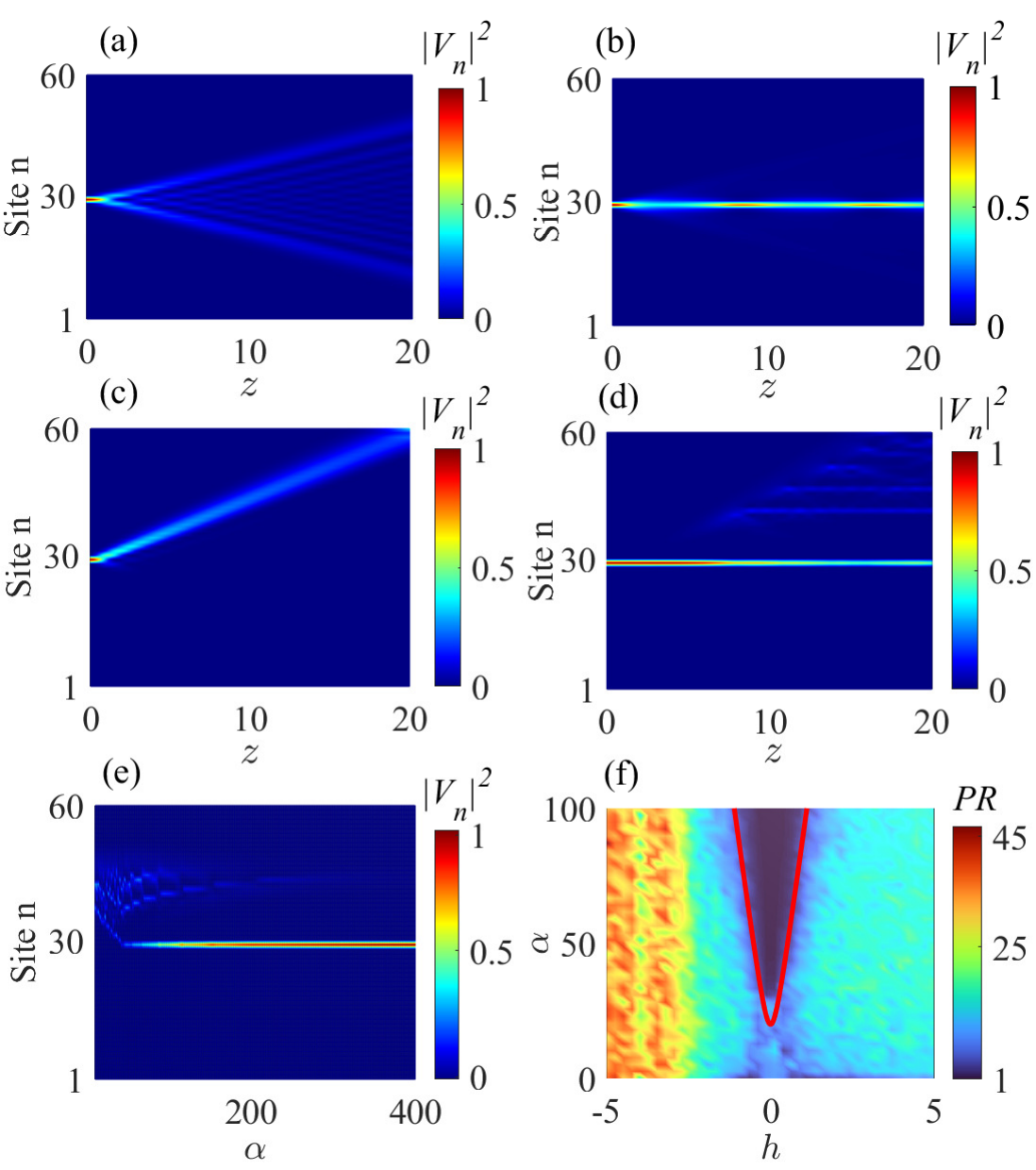}
		\end{center}
		\caption{Solitons in WGAs (with $N=60$), produced by Eq.~(\protect\ref{V}) with the single-channel excitation. Panels (a) and (b) correspond to the Hermitian DNLS equation with $C_{L}=C_{R}=1/2$: (a) the linear system, with $\protect\alpha =0$, and (b) the soliton generation triggered by ONC $\protect\alpha =2.25$. Panels (c) and (d) represent the non-Hermitian NH model, i.e., Eq.~(\protect\ref{V}), with $C_{L}=1/2$, $C_{R}=1$, and the same values of $\protect\alpha $ as in panels (a) and (b), respectively. (e) The output intensity distributions at $z=10$ vs. $\protect\alpha $, with $C_{L}=1/2$, $C_{R}=1$. (f) The PR of the output beam at $z=10$, defined as per Eq.~(\protect\ref{PR}), as a function of $\protect\alpha $ and $h$ with fixed $C_{L}=2$. The red curve is plotted according to Eq.~(\protect\ref{eq2}). }
		\label{fig2}
	\end{figure}
	
	\begin{figure}[th]
		\begin{center}
			\includegraphics[width=7.5cm]		{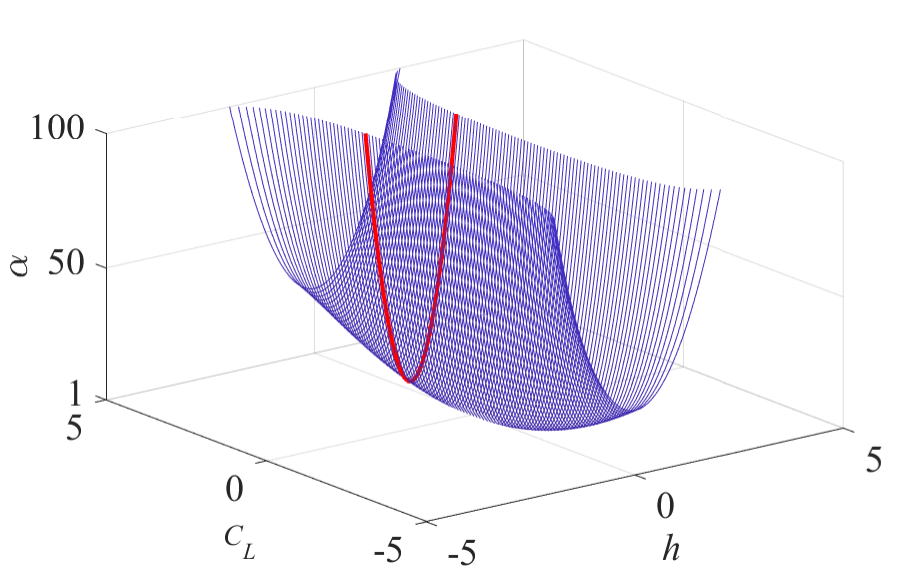}
		\end{center}
		\caption{The existence threshold for the lattice solitons in the nonlinear non-Hermitian WGAs under the single-site excitation. The purple surface is produced by Eq.~(\protect\ref{eq2}), which governs the interplay of the diffraction, nonlinearity, and non-Hermiticity. The surface partitions the parameter space into distinct regions, in which lattice solitons exist or not. The red curve corresponds to the one plotted in Fig.~\protect\ref{fig2}
			(f). }
		\label{fig3}
	\end{figure}
	
	In the linear limit, Eq.~(\ref{V}) with $\alpha =0$ admits an analytical solution under the initial condition Eq.~(\ref{input}),
	\begin{equation}
		U_{n}=A\left( \frac{-iC_{R}}{\sqrt{C_{R}C_{L}}}\right) ^{n}J_{n}\!\left( 2
		\sqrt{C_{R}C_{L}}z\right) ,  \label{Bessel}
	\end{equation}
	where $J_{n}$ is the Bessel function of order $n$ \cite{hou_controlling_2025}. This expression indicates that the spatial shape of the mode excited by the single-site input is the Bessel function of the propagation distance, exponentially modulated by {a} factor $\left( -iC_{R}/\sqrt{C_{R}C_{L}}\right) ^{n}$, as a function of the discrete coordinate $n$. Therefore, in the case of $C_{L}>C_{R}$, i.e., $h<0$ in Eq.~(\ref{h}), the mode exponentially decays in the direction of increasing $n$, and in the opposite direction in the case of $h>0$. As seen in Fig.~\ref{fig2}(c), the diffracted light converges toward the WGAs edge in the case of $C_{L}<C_{R}$, which may be considered as a signature of NHSE. The set of Figs.~\ref{fig2}(a)-(d) display four different propagation pictures, in the Hermitian [Figs.~\ref{fig2}(a)-\ref{fig2}(b)] and non-Hermitian [Figs.~\ref{fig2}(c)-\ref{fig2}(d)] cases, which demonstrate the effect of the nonlinearity. In particular, the Hermitian diffraction transport, governed by Eq.~(\ref{Bessel}) in Fig.~\ref{fig2}(a), evolves into a soliton regime upon the introduction of the nonlinearity, as seen in Fig.~\ref{fig2}(b).
	
	As for the non-Hermitian case, presented in Figs.~\ref{fig2}(c) and \ref{fig2}(d), the comparison with its Hermitian counterparts in Figs.~\ref{fig2}(a) and \ref{fig2}(b) reveals that the NHSE pulls the {beams} from bottom to top. The Kerr nonlinearity, represented by ONC $\alpha $, is the crucial factor in shaping the emergent soliton propagation. Specifically, in Fig.~\ref{fig2}(c) the diffracted light {becomes fully} localized under the action of NHSE. When $
	\alpha $ increases, the soliton regime emerges, although a part of the power may still diffract into the free space, as seen in Fig.~\ref{fig2}(d).
	
	We further analyze how ONC $\alpha $ {affects} the soliton generation efficiency in Figs.~\ref{fig2}(e), where $\alpha $ continuously increases from $0$ to $400$, and the snapshot of the light-intensity distribution is taken at a fixed propagation distance, $z=10$. It is found that, as $\alpha $ increases, a well-defined stable soliton forms, accompanied by a minimal energy loss. To explore the existence regions of the solitons, in Fig.~\ref{fig2}(f) we plot the PR of the output field at $z=10$, which is defined as
	\begin{equation}
		\text{PR}\left( h,\alpha \right) =\frac{\sum_{n}|U_n|^{4}}{
			(\sum_{n}|U_n|^{2})^{2}},  \label{PR}
	\end{equation}
	in the plane of the non-Hermiticity parameter $h$ (see Eq. (\ref{h})) and ONC $
	\alpha $. PR quantifies the spatial extent of the wavefunction, with smaller PR values indicating stronger localization. In Fig.~\ref{fig2}(f), low PR values are represented by deep-blue regions, where clean robust solitons emerge.
	
	To {further characterize} the relation between ONC $\alpha $ and the non-Hermiticity parameter $h$, we performed an extensive numerical analysis across a broad parameter space, spanning $C_{L,R}\in \lbrack 0,10]$ and $
	\alpha \in \lbrack 0,100]$. This process involved generating thousands of datasets similar to the results shown in Fig.~\ref{fig2}(f). By employing the automated algorithmic annotation technique, we extracted a vast collection of critical boundaries that delineate the transition between soliton-bearing and non-soliton regimes, which is conceptually similar to the red threshold line depicted in Fig.~\ref{fig2}(f). Subsequently, these data were processed using the symbolic-{regression} method, which is a robust data-driven machine-learning framework \cite{hafner_machine_2023}, to distill a semi-analytical expression that effectively identifies the operational boundary of the system:
	\begin{equation}
		3\left( 31-C_{L}^{2}\right) h^{2}+C_{L}^{2}-\alpha =0.  \label{eq2}
	\end{equation}
	The reasonable agreement between the red curves produced by Eq.~(\ref{eq2}) and the actual soliton/non-soliton threshold, which is observed in Fig.~\ref{fig2}(f), demonstrates the relevance of the prediction. Finally, Fig.~\ref{fig3} shows the overall boundary between the soliton-bearing and no-soliton behavior across all possible parameter settings.
	
	Equation~(\ref{eq2}), along with the corresponding plot in Fig.~\ref{fig3}, represents the results of the interplay of the non-Hermiticity, nonlinearity, and discrete diffraction in the optical WGA system. The general conclusions for the case of the single-channel excitation are that, to generate solitons in the WGA, the nonlinearity (or the input intensity) should be enhanced, while the non-Hermiticity acts on the light beams as an effective transverse {force} \cite{gill_atmosphere_ocean_1982}. Its enhancement hinders the soliton generation, making it necessary to resort to a stronger nonlinearity and/or larger input intensity to achieve the creation of the solitons.
	
	{\color{black}
		\subsection{Lattice solitons produced by the broad-pulse excitations}
		
		This subsection {focuses on} wave packets which span multiple lattice sites. If a wave packet (e.g., a spatial optical soliton) encompasses many sites, the variation of the field amplitude between adjacent sites is smooth, making it possible to introduce the continuum approximation, in which the discrete lattice coordinate $n$ is handled as a continuous one. In the framework of the discrete HN model (\ref{eq1}), we separate the Hamiltonian and non-Hamiltonian parts, setting
		\begin{equation}
			C_{L}=C-\frac{1}{2}h,C_{R}=C+\frac{1}{2}h,  \label{C}
		\end{equation}
		where $h$ is {the} non-Hermiticity parameter (\ref{h}), and introduce the continuum approximation by means of the Taylor expansion,
		\begin{equation}
			U_{n\pm 1}\approx U(n)\pm \frac{\partial U}{\partial n}+\frac{1}{2}\frac{
				\partial ^{2}U}{\partial n^{2}}.  \label{Taylor}
		\end{equation}
		Here, the first-order derivative term $\partial U/\partial n$ captures the asymmetry or drift of the wave packet, serving as the fundamental mechanism through which the non-Hermiticity $h$ exerts its influence, {while} the term $
		\partial ^{2}U/\partial n^{2}$ accounts for the diffraction arising from the discrete lattice structure. The result is the continuum NLS equation with the additional non-Hamiltonian term $\sim \partial U/\partial n$:
		\begin{equation}
			i\frac{\partial U}{\partial z}=C\frac{\partial ^{2}U}{\partial n^{2}}+h\frac{
				\partial U}{\partial n}+\sigma \left\vert U\right\vert ^{2}U.  \label{NLS1}
		\end{equation}
		To eliminate the trivial $2CU$ in the course of the substitution, we have used an additional substitution, $U(n,z)\equiv \tilde{U}\left( n,z\right) \exp \left( -2iCz\right) $, and {drop} the tilde.
		
		Finally, setting
		\begin{equation}
			n\equiv \sqrt{2C}x,~U\equiv \frac{u^{\ast }\left( x,z\right) }{\sqrt{|\sigma |}},~\epsilon \equiv -\frac{h}{\sqrt{2C}},  \label{scaling}
		\end{equation}
		transforms Eq. (\ref{NLS1}) into the standard NLS equation:
		\begin{equation}
			i\frac{\partial u}{\partial z}=-\frac{1}{2}\frac{\partial ^{2}u}{\partial x^{2}}-\epsilon \frac{\partial u}{\partial x}-\mathrm{sgn}(\sigma )\cdot \left\vert u\right\vert ^{2}u.  \label{NLS2}
		\end{equation}
		By means of additional rescaling, $x\equiv \epsilon ^{-1}\tilde{x}$, $
		u\equiv \epsilon \tilde{u}$, $z\equiv \epsilon ^{-2}\tilde{z}$, one can set $
		\epsilon \equiv 1$, but it may be more convenient to keep $\epsilon $ as a free parameter$,$ aiming to develop the perturbation theory.
		
		In the {zeroth-order} approximation, $\epsilon =0$, the commonly known soliton solution of Eq.~(\ref{NLS2}) with $\mathrm{sgn}(\sigma )=+1$, is
		\begin{equation}
			u_{\mathrm{sol}}=\eta \mathrm{sech}\left( \eta \left( x-cz\right) \right)\exp 
			\left[ icx+\frac{i}{2}\left( \eta ^{2}-c^{2}\right) z\right] ,  \label{sol}
		\end{equation}
		where $\eta $ is the amplitude, and $c$ is the ``velocity" (actually, it is the spatial tilt in the WGA). Two elementary dynamical invariants (conserved quantities) of the NLS equation with $\epsilon =0$ are the power and momentum,
		\begin{equation}
			P=\int_{-\infty }^{+\infty }\left\vert u(x)\right\vert ^{2}dx,~M=i\int_{-\infty }^{+\infty }u\frac{\partial u^{\ast }}{\partial x} dx.  \label{PM}
		\end{equation}
		For the soliton solution (\ref{sol}), the power and momentum are
		\begin{equation}
			P_{\mathrm{sol}}=2\eta ,~M_{\mathrm{sol}}=2c\eta .  \label{PMsol}
		\end{equation}
		
		In the presence of the perturbation term $-\epsilon \partial u/\partial x$ in Eq.~(\ref{NLS2}), the exact evolution equations for the power and momentum are
		\begin{equation}
			\frac{dP}{dz}=2\epsilon \mathrm{Im}\left( \int_{-\infty }^{+\infty }u\frac{
				\partial u^{\ast }}{\partial x}dx\right) ,~\frac{dM}{dz}=-2\epsilon \int_{-\infty }^{+\infty }\left\vert \frac{\partial u}{\partial x}
			\right\vert ^{2}dx.  \label{d/dz}
		\end{equation}
		The substitution of the zeroth-order approximation (\ref{sol}) and (\ref
		{PMsol}) in the evolution equations (\ref{d/dz}) yields simple equations:
		\begin{equation}
			\frac{d\eta }{dz}=-2\epsilon c\eta ,~\frac{dc}{dz}=-\frac{2}{3}\epsilon \eta ^{2}.  \label{ec}
		\end{equation}
		Starting with the initial condition for the quiescent soliton,
		\begin{equation}
			\eta (z=0)=\eta _{0},~c(z=0)=0,  \label{z=0}
		\end{equation}
		a solution of Eqs.~(\ref{ec}) can be cast in the form of
		\begin{equation}
			c(z)=-\frac{\eta }{\sqrt{3}}\tan \left( \frac{2}{\sqrt{3}}\epsilon \eta _{0}z\right) ,  \label{c}
		\end{equation}
		
		\begin{equation}
			\eta ^{2}(z)=\frac{\eta _{0}^{2}}{\cos ^{2}\left[ \frac{2}{\sqrt{3}}\epsilon \eta _{0}z\right] }.  \label{eta}
		\end{equation}
		It is seen that $\eta (z)$ diverges after a propagation distance of $z_{\mathrm{blowup}}=\sqrt{3}\pi /\left( 4\epsilon \eta _{0}\right) $. Further, upon integrating the velocity given in Eq.~(\ref{c}), we find the trajectory deviation of the perturbed soliton:
		\begin{equation}
			x=\frac{\ln [{\cos {(\frac{2\epsilon \eta _{0}z}{\sqrt{3}})}}]}{2\epsilon }. \label{local}
		\end{equation}
		\begin{figure}[th]
			\begin{center}
				\includegraphics[width=8.2cm]		{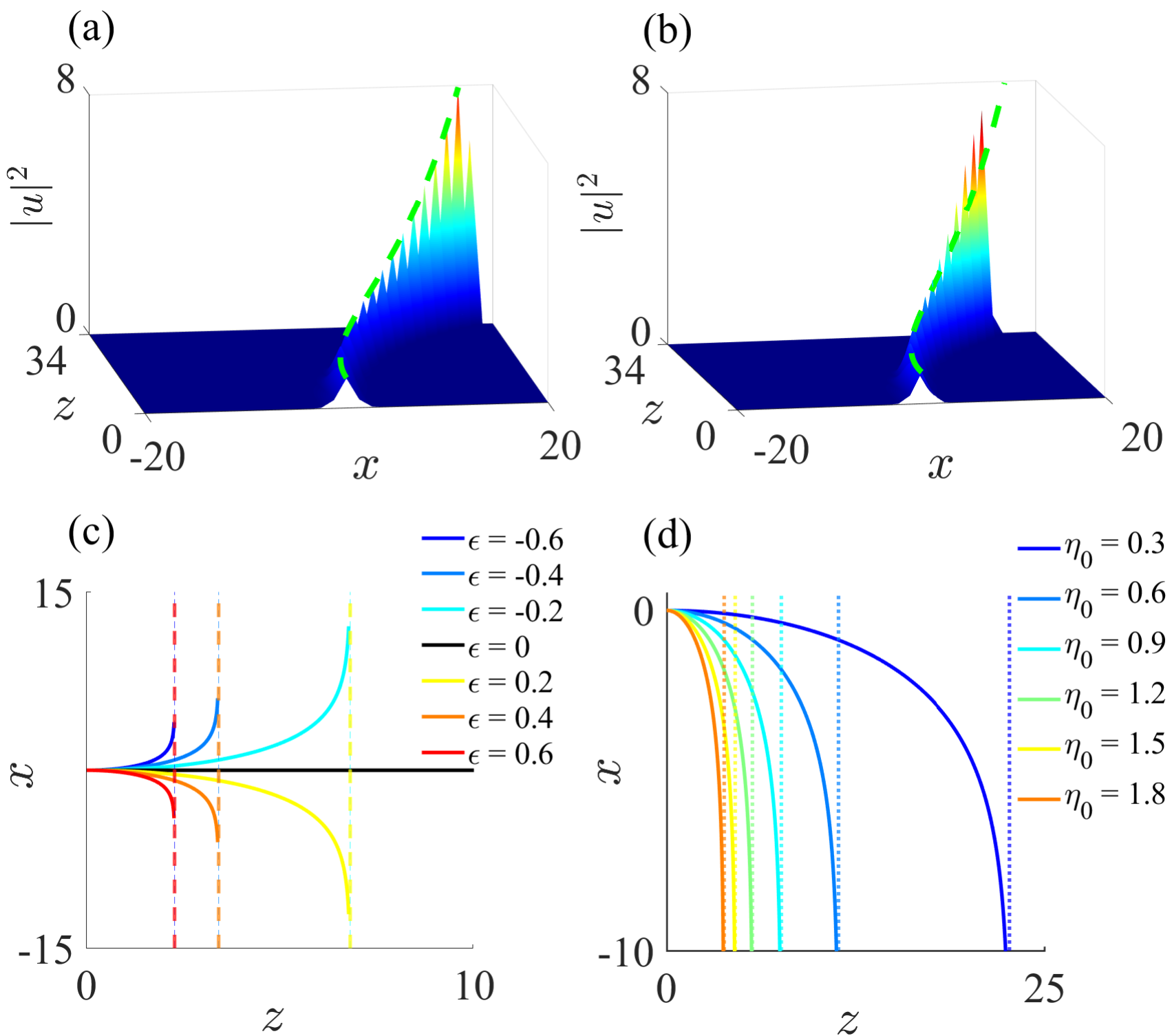}
			\end{center}
			\caption{\textcolor{black}{The soliton dynamics and boundary accumulation (NHSE) in the WGAs: (a) The simulation of the continuum-approximation NLS equation (\protect\ref
					{NLS2}). (b) The simulation of the underlying discrete HN-model equation~(\protect\ref{eq1}), where the green curves represent the soliton's amplitude and position defined as per Eqs.~(\protect\ref{eta}) and (\protect\ref{local}). Common parameters are $C=5$, $h=0.1$, $\protect\epsilon =-h/\protect\sqrt{
						2C}\approx -0.032$, and $\protect\eta _{0}=1$. (c) Soliton trajectories obtained by varying the perturbation strength $\protect\epsilon $ while keeping $\protect\eta _{0}=1$ constant. (d) Soliton trajectories obtained by varying the initial amplitude $\protect\eta _{0}$ while keeping $\protect%
					\epsilon =0.2$ constant. }}
			\label{fig4_add}
		\end{figure}
		
		Figure~\ref{fig4_add} {compares} the dynamical behavior produced by the numerical simulations of the continuum approximation, Eq.~(\ref{NLS2}), and the underlying discrete HN model~(\ref{eq1}) for broad-envelope initial excitation. For this purpose, Eq.~(\ref{NLS2}) was solved with input $u_{0}=\eta _{0}\mathrm{sech}(\eta _{0}x)e^{icx}$, while the initial condition for the discrete equation~(\ref{eq1}) was taken as $
		U=u_{0}^{\ast }/\sqrt{|\sigma |}$, with $x=n/\sqrt{2C}$. The simulations reveal that the pulse in the WGA accelerates towards right during the propagation, undergoing the continuous energy gain, and ultimately collides with the boundary. This phenomenon originates from the effect of the NHSE with $\epsilon <0$.
		
		The comparison of the plots in Figs.~\ref{fig4_add}(a) and \ref{fig4_add}(b) shows nearly identical evolution at the early stage, as produced by the continuum limit and the underlying HN model.\ Furthermore, the soliton's spatial trajectory and amplitude evolution, predicted by the perturbation theory [Eqs.~(\ref
		{local}) and (\ref{eta}), respectively], are superimposed, in the form of dashed green curves, on top of Figs.~\ref{fig4_add}(a) and \ref{fig4_add}
		(b). The results demonstrate excellent agreement in both cases, thereby validating the perturbation analysis.
		
		The reliability of the analytical {predictions} from perturbation theory makes it possible to systematically investigate the impact of the input's shape and non-Hermiticity on the propagation of broad pulses, using Eq.~(\ref{local}). With the input amplitude fixed as $\eta _{0}=1$, we have produced the pulse's propagation paths, varying the non-Hermiticity parameter $\epsilon $, as plotted in Fig.~\ref{fig4_add}(c). Further, fixing $\epsilon =0.2$, we examined the effect of varying the input-pulse's amplitude $\eta _{0}$, as shown in Fig.~\ref{fig4_add}(d). The findings clearly indicate that the action of the above-mentioned \textquotedblleft {effective transverse force}" $dc/dz$, induced by the non-Hermiticity as per Eq.~(\ref{ec}), {drives the soliton dynamics}. Naturally, the force grows with the increase of $\eta _{0}$, leading to a faster collision with the boundary.
	}
	\section{Nonlinear Hermitian bulk solitons and non-Hermitian skin solitons in nonlinear WGAs}
	\begin{figure}[th]
		\begin{center}
			\includegraphics[width=8cm]		{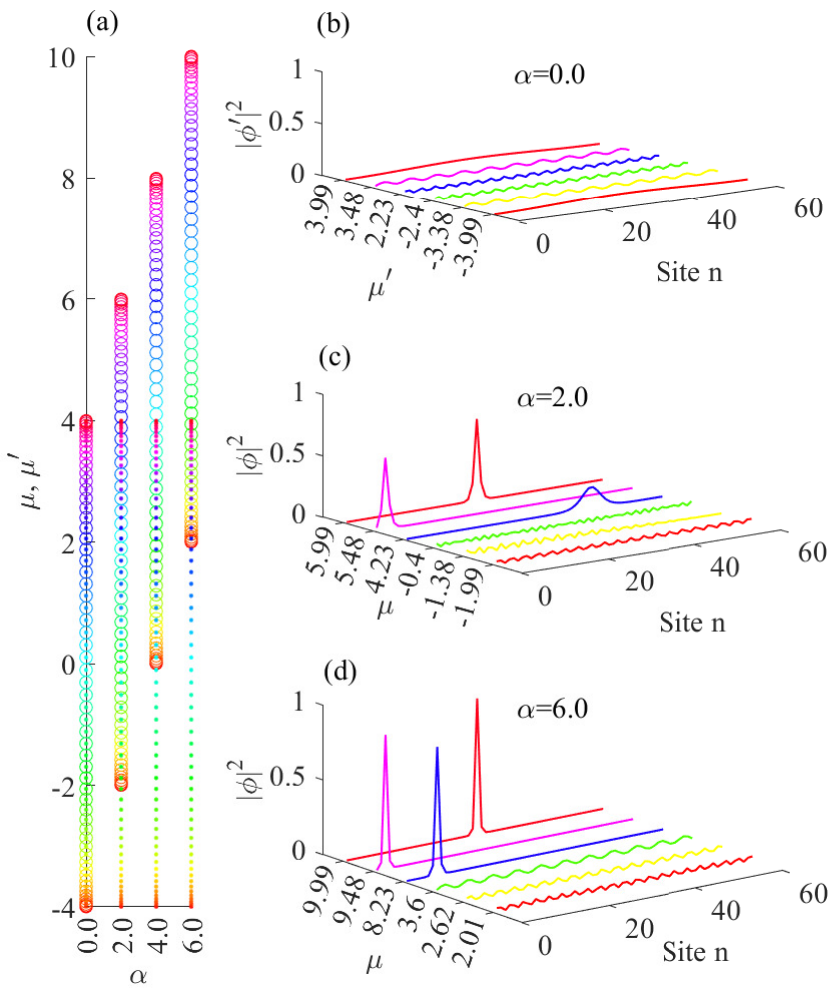}
		\end{center}
		\caption{
			Hermitian nonlinear propagation constant $\mu$ and nonlinear bulk modes obtained from the DNLS equation. (a) The propagation constants $\protect\mu ^{\prime }$ of the linear system (
			\protect\ref{eq11}) are marked by colored dots, while the spectrum of $\protect
			\mu $ for the full nonlinear system is represented by empty circles, varying the nonlinearity strength, $\protect\alpha =0,2,4,6$. (b) Selected six profiles of the linear modes $\protect\phi _{n}^{\prime }$. (c) and (d) Profiles of the nonlinear bulk modes $\protect\phi _{n}$. In panels (b)-(d), the eigenvalues $\protect\mu ^{\prime }$ or $\protect
			\mu $~are taken from panel (a) for the respective values of $\protect\alpha $ with the same color. Identical colors indicate that the linear solution $(\protect\mu ^{\prime },\protect\phi _{n}^{\prime })$ serves as the initial guess for finding the corresponding full nonlinear solutions $(\protect\mu ,\protect\phi _{n})$ by means of the Newton--Raphson method. The common parameters are $C_{L}=C_{R}=2,N=60$. }
		\label{fig5}
	\end{figure}
	
	\label{Sec4} Another relevant objective of the analysis of the nonlinear HN system is to identify skin and bulk modes \cite{manda_nonlinear_2025}, similar to the skin and bulk modes in the linear case \cite{Yao18}. To this end, we look for solutions to Eq.~(\ref{V}) in the usual form, $V_{n}=\phi _{n}e^{-i\mu z}$, with a stationary profile $\left\{ \phi _{n}\right\} $ and propagation constant $\mu $. Substituting the ansatz in Eq.~(\ref{eq1}) leads to a system of nonlinear algebraic equations,
	\begin{equation}
		C_{L}\phi _{n+1}+C_{R}\phi _{n-1}+\alpha |\phi _{n}|^{2}\phi _{n}=\mu \phi _{n}.  \label{eq10}
	\end{equation}
	We solved this nonlinear eigenvalue problem in two steps. First, the linear part of Eq.~(\ref{eq10}) is considered,
	\begin{equation}
		C_{L}\phi _{n+1}^{\prime }+C_{R}\phi _{n-1}^{\prime }=\mu ^{\prime }\phi _{n}^{\prime },  \label{eq11}
	\end{equation}
	solving it for eigenvalue $\mu ^{\prime }$ and eigenstate $\left\{ \phi _{n}^{\prime }\right\} $. At the second step, we use the solution of the linear system (\ref{eq11}) as an input (initial guess) for the solution of the full nonlinear equation~(\ref{eq10}) by means of the Newton-Raphson method. So obtained numerical solutions are presented in Figs.~\ref{fig5} and \ref{fig6} for the Hermitian and non-Hermitian cases, respectively.
	
	\begin{figure}[th]
		\begin{center}
			\includegraphics[width=8.7cm]	{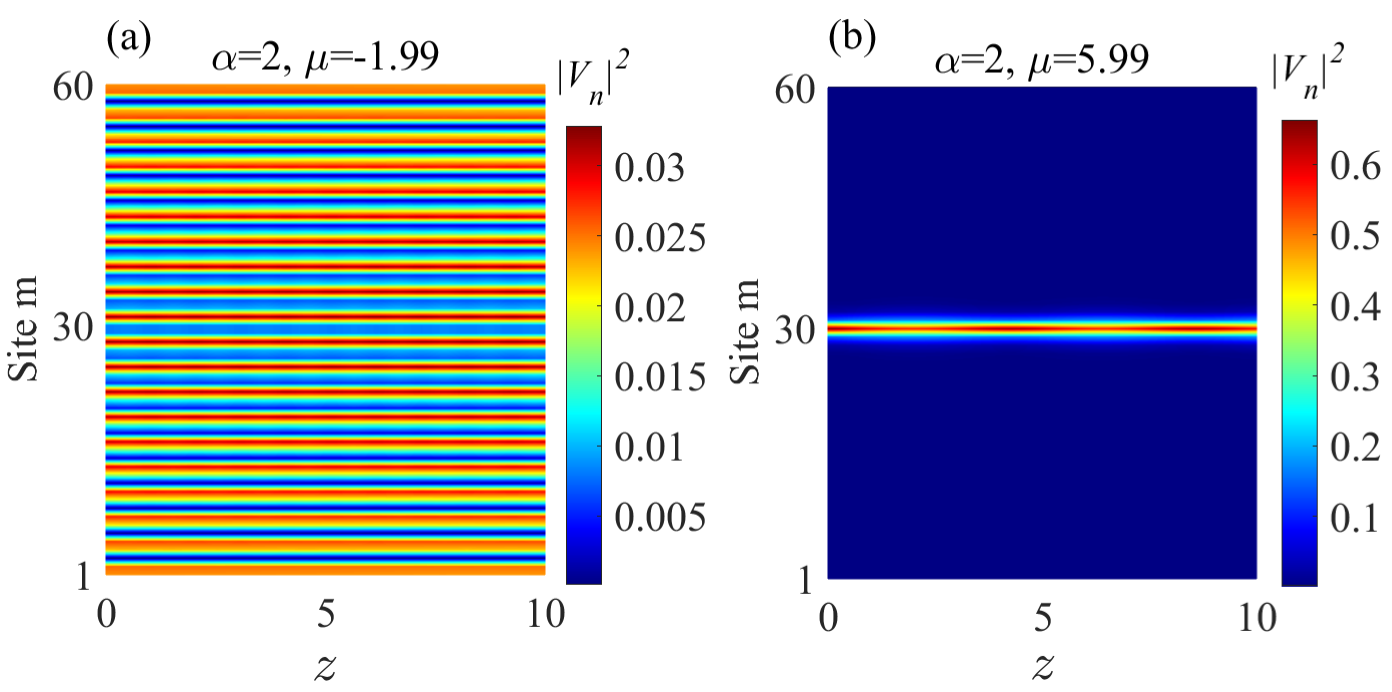}
		\end{center}
		\caption{  
			Dynamical simulations of light propagation from $z=0$ to $10$, using four distinct nonlinear eigenmodes as initial conditions. The titles of subfigures (a)–(b) indicate the simulation parameters: the nonlinearity strength $\alpha$ and the nonlinear propagation constant $\mu$, corresponding to those shown in Fig.~\ref{fig5}. The simulations verify the nonlinear eigenmodes, which retain their stationary profiles during propagation.}
		\label{fig7_1}
	\end{figure}
	
	Figure~\ref{fig5}(a) displays the Hermitian spectra of eigenvalues {for} four distinct values of ONC $\alpha $. For comparison, the Hermitian linear eigenvalues $\mu ^{\prime }$ (colored dots) and full nonlinear ones $\mu $ (colored empty circles) are superimposed for each value of $\alpha $. The identical color coding for the $\mu ^{\prime }$ dot and $\mu $ circle pertaining to each value of $\alpha $ implies that the corresponding Hermitian linear eigenstate ($\mu ^{\prime },\phi ^{\prime }$) served as the input for the generation of the respective nonlinear solution (
	$\mu ,\phi $). A noticeable difference between $\mu $ and $\mu ^{\prime }$ is produced by the Kerr nonlinearity. 
	
	Figure~\ref{fig5}(b) shows six representative eigenmodes selected from the complete set of sixty Hermitian bulk eigenmodes, which also serve as the input for generating the nonlinear solutions presented in Figs.~\ref{fig5}(c) and \ref{fig5}(d). For the linear ($\alpha =0$) and nonlinear ($\alpha =2$ and $6$) Hermitian lattice ($C_{L}=C_{R}\equiv 2$), these modes reveal that the nonlinear eigenvalue problem exhibits highly complex dynamics. The numerically obtained nonlinear eigenstates can be categorized into two distinct dynamical regimes according to the relative magnitude of their eigenvalues. For small eigenvalues, the Hermitian nonlinear modes are essentially uniform, resembling sine or plane waves, as  the array fluctuations are minimal. For large eigenvalues, solitons appear: they show no boundary preference and stay distributed across the lattice interior when considering all sixty nonlinear modes (only six are shown here due to space limitations), hence we refer to these nonlinear bulk modes as bulk solitons. Fig.~\ref{fig7_1} {shows the dynamics of Eq.~(\ref{V}) simulated using} two distinct eigenmodes (with nonlinear eigenvalues $\mu=-1.99$ and $\mu=5.99$, as identified in Figs.~\ref{fig5}(c)) as initial conditions. These results confirm that our numerical algorithm yields accurate eigenvalues and eigenstates, as they remain independent of $z$. Interestingly, the nonlinear bulk mode in Fig.~\ref{fig7_1}(b) exhibits behavior similar to the soliton dynamics simulated from the single-channel initial condition in Fig.~\ref{fig2}(b).
	
	The Hermitian version of Eq.~(\ref{eq10}) is
	\begin{equation}
		C\left( \phi _{n+1}+\phi _{n-1}\right) +\alpha |\phi _{n}|^{2}\phi _{n}=\mu \phi _{n}.  \label{Hermitian}
	\end{equation}
	In the linear limit, $\alpha =0$, the set of obvious exact solutions to Eq.~(
	\ref{Hermitian}), which satisfy the zero boundary conditions  at $n=0$ and $
	n=N+1$, is
	\begin{eqnarray}
		\phi _{n}^\prime &=&\Phi _{0}\sin (kn), \label{phi}\\
		\mu^\prime  &=&2C\cos k,  \label{mu}
	\end{eqnarray}
	where $\Phi _{0}$ is an arbitrary amplitude, and $k=\frac{m\pi }{N+1},m=1,2,\dots,N$. Then, 
	It is straightforward to verify that the full nonlinear equation (\ref{Hermitian}) supports {plane-wave eigenstates} of the form  $\phi_{n}=e^{ikn}$ (in unnormalized form, with the normalization factor given by $1/\sqrt{N}$)  with eigenvalue $\mu = \mu^\prime+\alpha$. This conclusion is corroborated by Fig.~\ref{fig5}, where the eigenvalues match those in Fig.~\ref{fig5}(a) exactly. Furthermore, the nonlinear eigenstates shown in Figs.~\ref{fig5}(c)–\ref{fig5}(d) approximate plane waves in {the} small-eigenvalue regime. Investigation of the bulk soliton regimes is reserved for future work.
	
	\begin{figure}[th]
		\begin{center}
			\includegraphics[width=8.2cm]		{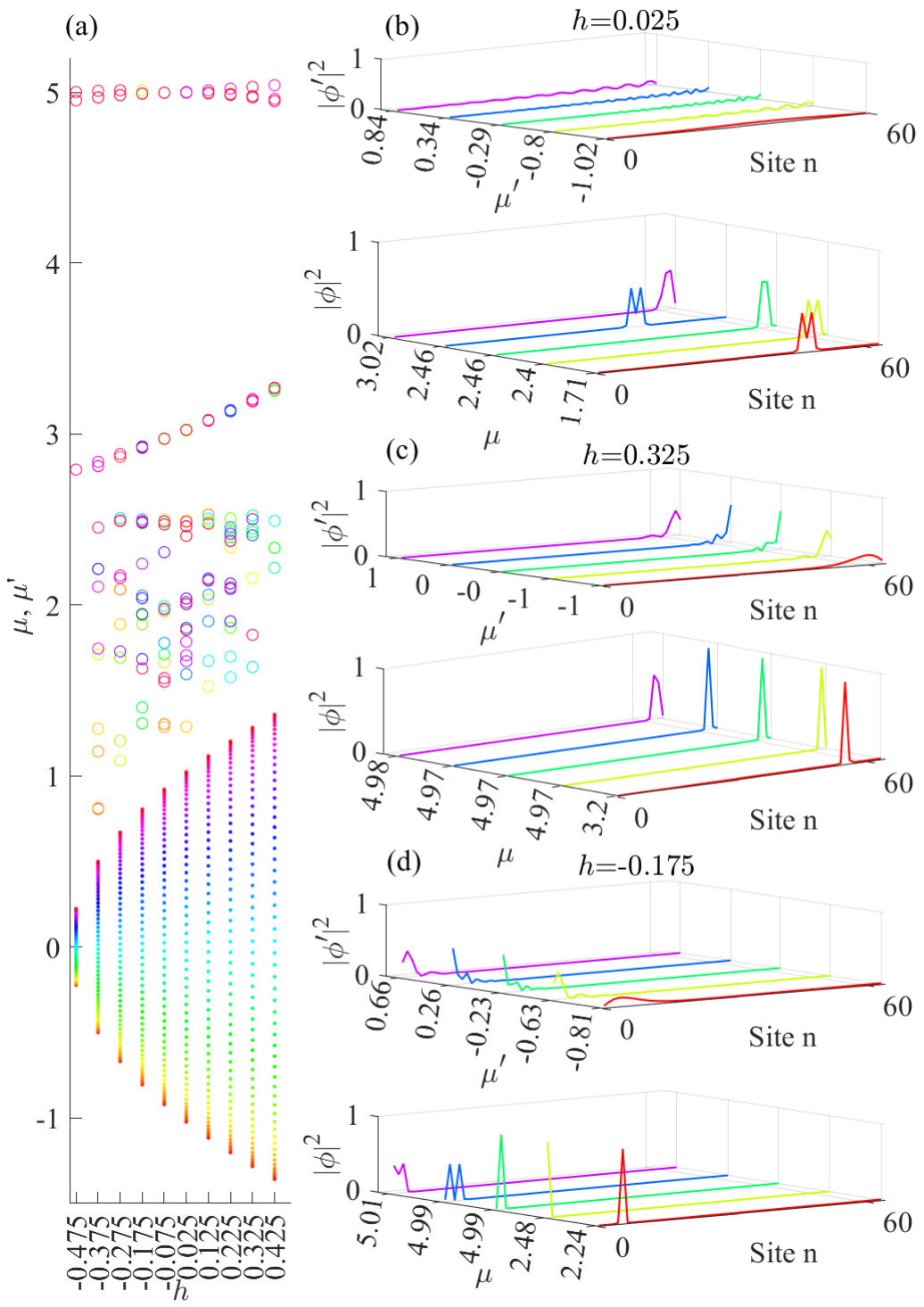}
		\end{center}
		\caption{The nonlinear HN skin solitons and their energy spectrum. Similar to Fig.~\protect\ref{fig5}, panel (a) shows the energy spectra of the linear system ($\protect\mu ^{\prime }$) and full nonlinear one ($\protect\mu $) as functions of non-Hermiticity $h$ (see Eq.~(\protect\ref{h})), with~$C_{R}$ ~ranging from $1/40$ to $1$, i.e., $h$ from $-19/40$ to $1/2$. Panels (b)-(d) present the comparison of three sets of the eigenstates for different values of $h$. In each group, the top subplots in the first row show linear skin modes, and the second row shows their nonlinear counterparts. The parameters are $\protect\alpha =5$, $C_{L}=1/2$, $N=60$. }
		\label{fig6}
	\end{figure}

	Fig.~\ref{fig6} shows the nonlinear energy spectrum and eigenstates of the non-Hermitian HN model, with eigenstates exhibiting more drastic modal changes than those of its Hermitian (DNLS) counterpart due to the NHSE. All nonlinear eigenstates tend to localize at the edges, forming what we term skin solitons. 
	Notably, the localization of the skin solitons {is} continuously tunable via the non-Hermiticity parameter $h$: the increase of $|h|$ systematically drives the wave packets from quasi-bulk distributions toward pronounced localization near the boundary. 
	As demonstrated in Figs.~\ref{fig6}(b)–\ref{fig6}(d), which show five selected modes out of sixty, the linear localized skin modes (top row) evolve into skin solitons (bottom row) as the non-Hermiticity parameter $h$ increases.
	{When $h$ is small, the localization of the skin solitons near the lattice edge is relatively weak.} With {the increase of }$h$, both the linear skin modes and skin solitons exhibit strong shrinkage toward the edge (see the right boundary in Figs.~\ref{fig6}(b) and \ref{fig6}(c), and the left one in Fig.~\ref{fig6}
	(d)). We refer to these nonlinear localized modes as skin solitons to distinguish them from the bulk modes observed in the Hermitian system, as their structure is determined by the combined effects of the NHSE and Kerr nonlinearity.
	
	The comparison between the nonlinear bulk modes [Figs.~\ref{fig5}(c) and \ref{fig5}(d)] and skin solitons [Figs.~\ref{fig6}(b)-\ref{fig6}(d)] reveals that NHSE remains a major factor in the nonlinear regime. It effectively  {compresses} the bulk modes onto the boundary, leading to the formation of {distinct} skin modes. Throughout this process, the localization of the standard linear skin modes is considerably reinforced, thereby confining the energy more effectively and rendering the propagation increasingly stable. This feature holds promise for enhancing the performance of topological lasers and various photonic applications. Theoretically, it is also of considerable significance for the investigation of nonlinear topological boundary states. Furthermore, the non-Hermitian-effect-induced boundary compression of the nonlinear bulk modes can be interpreted as a nonlinear enhancement of the NHSE, bearing qualitative similarity to its linear analogue \cite{Yao18}.
	
	\begin{figure}[th]
		\begin{center}
			\includegraphics[width=8cm]		{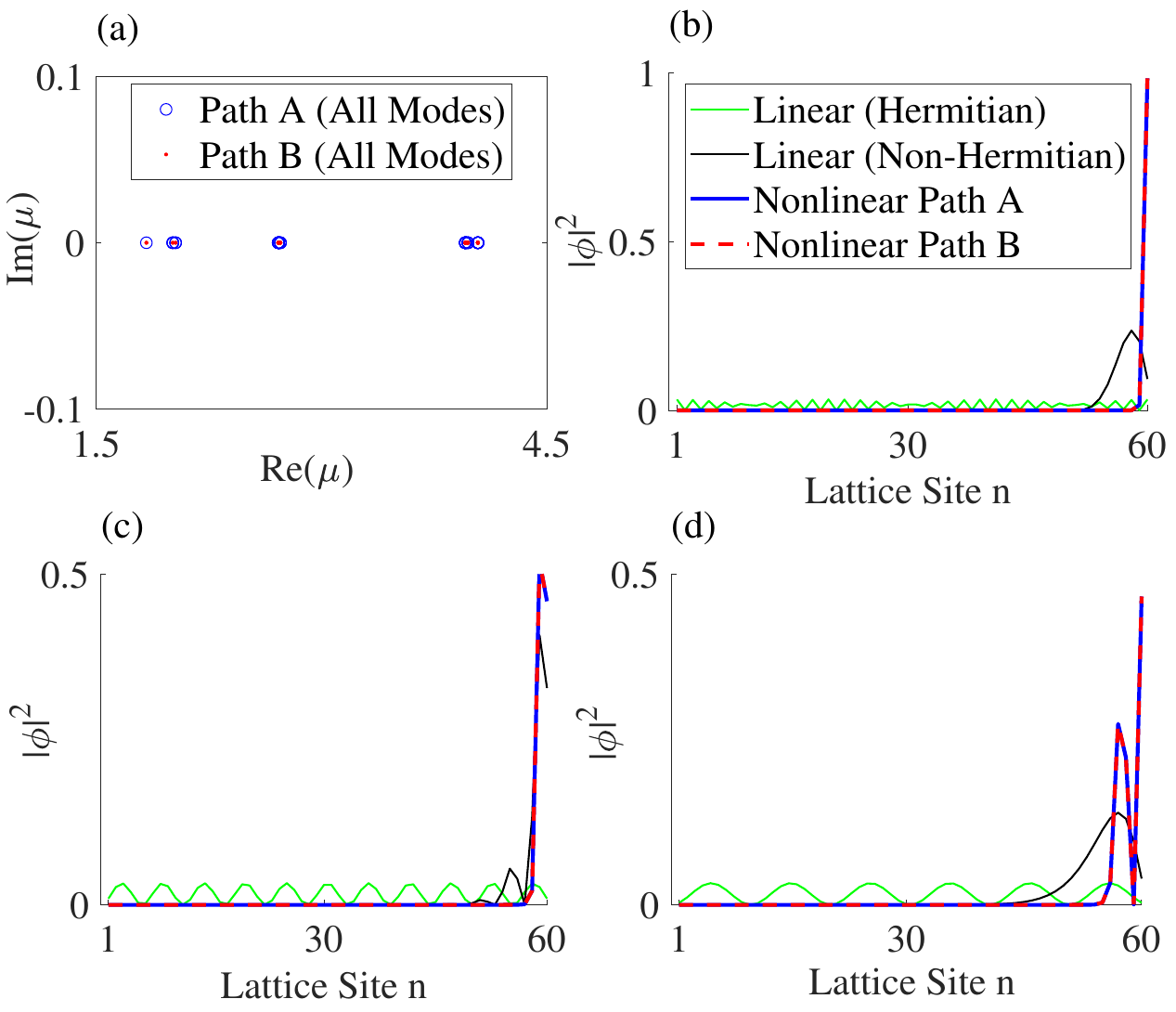}
		\end{center}
		\caption{{
				The comparison of the final energy spectra and selected eigenstates obtained from Paths A and B, demonstrating the robust one-to-one correspondence and path-independence for the nonlinear skin modes. (a) The perfect overlap of the complex energy spectra from Paths A (blue circles) and B (red dots) with degeneracy for certain modes. (b)-(d) Spatial profiles of three representative nonlinear non-Hermitian eigenstates with eigenvalues $\mu = 4.0399, 2.7130,$ and $1.8328$, respectively. Both paths precisely converge to identical final localized states (blue/red lines), regardless of their distinct linear initial states (green/black lines). Parameters for Path A: $\alpha=0, C_{L}=C_{R}=1/2$, followed by tuning $\alpha$ to 4, and then $C_{R}$ to $0.825$. Parameters for Path B: $\alpha=0, C_{L}=1/2, C_{R}=0.825$, followed by tuning nonlinearity $\alpha$ to $4$.}}
		\label{fig8}
	\end{figure}
	
	To further confirm the robustness of the connection between the linear and nonlinear regimes and rule out any potential path-dependence or hysteresis, we performed the following verification by tracking the states via two distinct pathways (A and B), as illustrated in Fig.~\ref{fig8}. In Path A, we start from a linear Hermitian lattice ($\alpha=0, C_{L}=C_{R}=1/2$), first introduce the nonlinearity to establish Hermitian nonlinear modes ($\alpha=4, C_{L}=C_{R}=1/2$), and subsequently introduce non-Hermiticity by tuning $C_{R}$ to the target value $0.825$. Conversely, in Path B, we initialize the system with linear non-Hermitian skin modes ($\alpha=0, C_{L}=1/2, C_{R}=0.825$) and then adiabatically increase the nonlinearity to the target value $\alpha=4$.
	Figures~\ref{fig8}(a) compare the final energy spectra {obtained from} the two distinct paths, showing agreement between them, with some modes exhibiting degeneracy. Figures~\ref{fig8}(b)-\ref{fig8}(d) present the comparison between the results for three randomly chosen different nonlinear non-Hermitian eigenfunctions at the end of the process performed along the two paths, with nonlinear eigenvalues $4.0399$, $2.7130$, and $1.8328$, respectively. The comparison confirms that the eigenstates are identical. The detailed spatial profiles of the representative modes [Figs.~\ref{fig8}(b)--\ref{fig8}(d)] further reveal that, despite starting from completely different linear initial states (represented by the green and black lines), both pathways precisely converge to exactly the same final nonlinear non-Hermitian localized states (represented by the blue and red lines).
	
	It should be noted that {while nonlinear systems can, in principle, support a vast diversity of solutions arising from complex bifurcations}, we focus on the nonlinear counterparts of the linear eigenmodes. 
	To this end, we construct these states by applying the Newton-Raphson method starting from 60 linear eigenstates as initial seeds. This approach ensures a one-to-one correspondence that reveals how the inherent lattice geometry and non-Hermiticity reshape the fundamental mode properties. The skin solitons constitute a physically significant well-defined family of states that directly originate from their linear topological counterparts.

	\section{Conclusion}\label{Sec5}
	This paper addresses light propagation phenomenon under the interplay of the NHSE and Kerr nonlinearity in the nonlinear HN optical WGAs with nonreciprocal coupling. The dynamics are examined under single-site and broad-pulse excitations. 
	The single-site case reveals a competitive mechanism for soliton generation: solitons emerge from the interplay between the NHSE and nonlinearity, wherein nonlinear effects counteract boundary localization while the NHSE drives power toward the edges.
	An empirical formula, derived by means of the data-driven symbolic regression method, accurately delineates the boundary between the soliton-bearing and non-soliton regimes. Under the broad-pulse excitation, the perturbative solution to the continuum approximation of the HN model reveals that the interplay between the nonlinearity and non-Hermiticity triggers rapid acceleration of the wave packet driven by an effective transverse force, along with significant amplification, leading to robust light localization at the WGA boundaries. 
	For the stationary problem, the numerical study for the nonlinear eigenvalue problem reveals that the nonlinearity transforms the linear modes into nonlinear bulk modes in the Hermitian WGA, whereas in the non-Hermitian case the modes are driven toward the edges, forming skin solitons attached to the WGA boundary. This phenomenon represents the nonlinear manifestation of the NHSE, effectively extending it from the linear regime to the nonlinear one.
	
	The findings above uncovering novel ``exotic" localized waves advance our understanding of light transport in WGAs or photonic lattices. Beyond the current findings, several promising research avenues remain open. For instance, one may ask whether nonlinear topological edge modes exist in traditional linear topological systems. Furthermore, extending this framework to higher-dimensional systems, such as two-dimensional arrays, could reveal a rich interplay between nonlinearity and higher-order topological phases.
	These results may not only contribute for the design of novel integrated photonic circuits, optical switches, and topological lasers, but also offer pioneering insights into the burgeoning field of nonlinear non-Hermitian physics.
	
	\begin{acknowledgments}
		This work was supported by the National Natural Science Foundation of China (Grant Nos. 11975172, 12261131495 and 12381240286).
	\end{acknowledgments}
	
	\bibliographystyle{myprb}
	\bibliography{bibfile}
	
\end{document}